\documentclass[runningheads]{llncs}
\usepackage[T1]{fontenc}
\usepackage{graphicx}
\usepackage{graphicx}
\usepackage{subcaption}
\usepackage{listings}
\usepackage{xcolor}
\usepackage{array} 
\usepackage[inline]{enumitem}
\usepackage{booktabs}
\usepackage{amssymb}

\makeatletter
\renewcommand\@fnsymbol[1]{}
\makeatother

\begin{document}
\title{Evaluating Multi-Hop Reasoning in RAG Systems: A Comparison of LLM-Based Retriever Evaluation Strategies
\thanks{Accepted for publication at the SynIRgy Workshop, ECIR 2026 (48th European Conference on Information Retrieval).}
}
\titlerunning{Multi-Hop Retriever Evaluation}

%
%
\author{Lorenz Brehme\orcidID{0009-0009-4711-2564} 
\and
Thomas Ströhle\orcidID{0000-0002-1954-6412} \and
Ruth Breu\orcidID{0000-0001-7093-4341}}
\authorrunning{Brehme et al.}

\institute{Universität Innsbruck, Technikerstra\ss e 21a, 6020 Innsbruck, Austria\\\email{\{lorenz.brehme,thomas.stroehle,ruth.breu\}@uibk.ac.at}}
\maketitle

\begin{abstract}

Retrieval-augmented generation (RAG) enhances large language models (LLMs) with external knowledge to answer questions more accurately.
However, research on evaluating
RAG systems—particularly the retriever component—remains limited, as most existing work focuses on single-context retrieval rather than multi-hop queries, where individual contexts may appear irrelevant in isolation but are essential when combined.
In this research, we use the HotPotQA, MuSiQue, and SQuAD datasets to simulate a RAG system and compare three LLM-as-judge evaluation strategies, including our proposed Context-Aware Retriever Evaluation (CARE). Our goal is to better understand how multi-hop reasoning can be most effectively evaluated in RAG systems.
Experiments with LLMs from OpenAI, Meta, and Google demonstrate that CARE consistently outperforms existing methods for evaluating multi-hop reasoning in RAG systems. The performance gains are most pronounced in models with larger parameter counts and longer context windows, while single-hop queries show minimal sensitivity to context-aware evaluation. Overall, the results highlight the critical role of context-aware evaluation in improving the reliability and accuracy of retrieval-augmented generation systems, particularly in complex query scenarios.
To ensure reproducibility, we provide the complete data of our experiments at \url{https://github.com/lorenzbrehme/CARE}.

\end{abstract}

\section{Introduction}

Retrieval-Augmented Generation (RAG) is an ongoing trend in research and has been continuously developed over the past years~\cite{brehme_retrieval-augmented_2025}. It augments their knowledge through the integration of external data sources and addresses hallucinations in large language models (LLMs) ~\cite{lewis_retrieval-augmented_2021}. A basic RAG system consists of three components~\cite{gao_retrieval-augmented_2024}:  \begin{enumerate*}[label=($\roman*$)]
\item Indexing component---responsible for indexing and embedding external data.
\item Retriever component---filters and retrieves relevant information based on the input query.
\item Generator component---uses an LLM to process the retrieved data together with the input query to generate the final response.
\end{enumerate*}
Evaluating all three components is essential for assessing the performance of a RAG system and ensuring it produces accurate, well-grounded responses: While the generator component has been extensively studied---with approaches such as RAGAS~\cite{es_ragas_2023} and Ares~\cite{saad-falcon_ares_2024} proposing methodologies to assess the quality of RAG-generated responses---evaluation of the retrieval and indexing components has received comparatively less attention~\cite{brehme2025SLR}. In particular, assessment of indexing methods often relies on system-level performance metrics such as throughput and latency~\cite{kukreja_performance_2024}.

Our focus is on retriever evaluation, which assesses the relevance of the context retrieved by a RAG system. This evaluation is typically conducted using one of two main strategies: labeled~\cite{moreira_enhancing_2024,tang_multihop-rag_2024} and unlabeled~\cite{afzal_towards_2024,salemi_evaluating_2024}. In labeled evaluations, the relevant context for each query is predefined within the dataset, and the retriever is assessed based on its ability to return these labeled documents~\cite{tang_multihop-rag_2024}. However, labeled evaluation is not always practical. When the chunking strategy is modified, the segments labeled as relevant may no longer correspond to the actual segments retrieved by the RAG system due to changes in document segmentation~\cite{liu_cofe-rag_2024}. Additionally, the structure of QA datasets poses challenges: each question-answer pair is typically derived from documents in the corpus and labeled accordingly~\cite{es_ragas_2023,tang_multihop-rag_2024}. This process does not guarantee that all other documents are irrelevant, as no negative sampling is applied~\cite{xu_negative_2022}. As a result, a retriever may return documents not labeled as relevant that still contain essential information for answering the query~\cite{brehme2025SLR}.

Therefore, unlabeled approaches have become prevalent, as they assess context relevance without requiring labeled documents~\cite{brehme2025SLR}. In this setting, only the input query and the ground-truth answer are used to determine whether the retrieved contexts are relevant.This assessment can be carried out by human annotators~\cite{afzal_towards_2024} or through automated approaches using LLMs, as demonstrated in evaluation frameworks such as the one introduced by ~\cite{rackauckas_evaluating_2024}.
In unlabeled evaluation approaches, existing methods, like direct~\cite{afzal_towards_2024,rackauckas_evaluating_2024,saad-falcon_ares_2024} and indirect evaluation~\cite{salemi_evaluating_2024} do not account for multi-hop queries. These methods typically evaluate each retrieved context in isolation, ignoring the relationships and dependencies among multiple retrieved documents. This limitation is particularly critical for multi-hop queries, which require integrating information from multiple sources to derive a correct answer. 

This work investigates effective evaluation of RAG systems in unlabeled settings, focusing on multi-hop queries. We compare three evaluation strategies in a synthetic RAG environment and benchmark them on HotPotQA, MuSiQue, and SQuAD using LLMs from OpenAI, Meta, and Google. We analyze how question difficulty and query type affect performance. Our results show that CARE outperforms existing approaches, with the largest gains on complex and comparative multi-hop questions—especially when using models with longer context windows, larger parameter counts, and shorter prompts. Notably, while single-hop queries generally do not require context awareness, multi-hop evaluation benefits substantially from it, underscoring the importance of context-aware evaluation for improving the reliability and accuracy of RAG systems in challenging scenarios.

\section{Related Work}

The use of LLMs for retriever evaluation has been explored in several studies~\cite{friel_ragbench_2024,saad-falcon_ares_2024,alinejad_evaluating_2024}. Typically, an LLM is prompted with an input query, an answer, and a retrieved context and is tasked with determining whether the context is relevant to answering the query~\cite{brehme2025SLR}. In the literature, two primary definitions of context relevance have emerged. The first definition characterizes context relevance based on whether the context was actually used by the LLM during answer generation~\cite{friel_ragbench_2024,ding_vera_2024}. The second definition considers a context relevant if the information retrieved is useful for answering the question~\cite{saad-falcon_ares_2024}. In this work, we adopt the second definition, as the former is not applicable to our setting: it relies on the internal behavior of the generator.
Existing work largely falls into two categories for measuring context relevance: \textit{direct} (cf. Figure \ref{fig:direct-model}) and \textit{indirect} (cf. Figure \ref{fig:indirect-model}) evaluation approaches~\cite{brehme2025SLR}.
In \textit{direct} evaluation, the LLM is provided with the input query, the ground-truth answer, and the context to be evaluated. The LLM is then tasked with determining, based on a specific prompt, whether the context is relevant to answering the question. 
Prior work has explored various labeling schemes under this paradigm, including graded relevance scales~\cite{rackauckas_evaluating_2024}, binary relevance classification~\cite{saad-falcon_ares_2024}, and sentence-level relevance estimation~\cite{es_ragas_2023}.
In contrast, the \textit{indirect} approach preprocesses the retrieved context and input query before assessing context relevance. For example, \cite{alinejad_evaluating_2024} generate an answer from the retrieved context list and compare it to an answer derived from a golden document list containing all relevant contexts. Contexts are considered relevant if the answers match. This method limits labeling to the entire context set, preventing individual relevance assignments. Another adaptation by \cite{salemi_evaluating_2024} generates an answer for each context individually and labels a context as relevant if the answer is correct.
While this method allows for individual labels for each context, it does not account for multi-hop queries, where the answer depends on multiple pieces of context.
Since multi-hop reasoning is a common requirement in RAG systems~\cite{tang_multihop-rag_2024}, our research investigates and compares different unlabeled, LLM-based evaluation approaches. Our goal is to identify effective strategies for retriever evaluation that capture the interdependencies among multiple retrieved contexts and to determine which evaluation methods best reflect retrieval quality in multi-hop settings.

\section{Datasets}
To simulate the RAG pipeline, each dataset must include a question, a set of retrieved documents containing both relevant and non-relevant contexts, and a ground-truth answer to support evaluation. The inclusion of labeled relevant and non-relevant contexts is essential, as it enables evaluation of whether the LLM correctly distinguishes relevant information from not relevant.
For our experiments, we simulate the RAG pipeline using the multi-hop datasets HotPotQA~\cite{yang2018hotpotqa} and MuSiQue~\cite{trivedi2021musique}, as well as the single-hop dataset SQuAD 2.0~\cite{rajpurkar_know_2018}.
While HotPotQA and MuSiQue explicitly include irrelevant contexts in their original structure, SQuAD 2.0 requires adaptation. In SQuAD 2.0, each context document is associated with a set of questions that are either answerable or unanswerable based on that specific text. To transform this into a retriever-style format, we treat the original context as relevant if it contains the answer and as non-relevant if it does not.
We then augment each question–context pair with additional distractor contexts. Specifically, we use the full SQuAD corpus to retrieve the top 19 most similar documents per question using BM25 \cite{trotman_improvements_2014}, resulting in a context list of 20 contexts for each query. These additional contexts serve as unlabeled context documents to simulate a retriever returning multiple documents, though only the originally labeled context is used for the final relevance evaluation.
While SQuAD 2.0 contains only single-hop queries, HotPotQA and MuSiQue are well suited for evaluating multi-hop reasoning, as they provide question–answer pairs with multiple context documents, typically two of which are labeled as relevant. For our experiments, we selected instances with exactly two relevant contexts and used the datasets as provided, enabling a consistent multi-hop evaluation. This yields context lists of 10 documents per question for HotPotQA and 20 for MuSiQue.

Further granularity is provided by the HotPotQA dataset, which categorizes questions into easy, medium, and hard difficulty levels. Easy questions are primarily single-hop, while medium and hard questions require multi-hop reasoning. The distinction between the latter two levels is based on the performance of state-of-the-art models as reported by \cite{yang2018hotpotqa}, where questions with a success rate below 60\% are classified as hard.
Additionally, the dataset includes two distinct question types, bridge and comparison, described by~\cite{yang2018hotpotqa}: Bridge questions require connecting multiple pieces of information from different contexts to arrive at the answer.
For example, in the question \textit{``Where was the singer and songwriter of Radiohead born?''}, the bridge entity is Radiohead. To answer this, one must first determine who the singer and songwriter of Radiohead is, and then find that person’s place of birth. 
Comparison questions, on the other hand, involve evaluating and comparing facts across multiple contexts. 
For instance, the question \textit{``Who has played more NBA games, Michael Jordan or Kobe Bryant?''} requires retrieving the number of NBA games each player has played from different sources, then comparing those values to find the answer.

\section{Evaluation Strategies}

\begin{figure}[htbp]
  \centering
  \begin{subfigure}[b]{0.3\textwidth}
    \includegraphics[width=\linewidth]{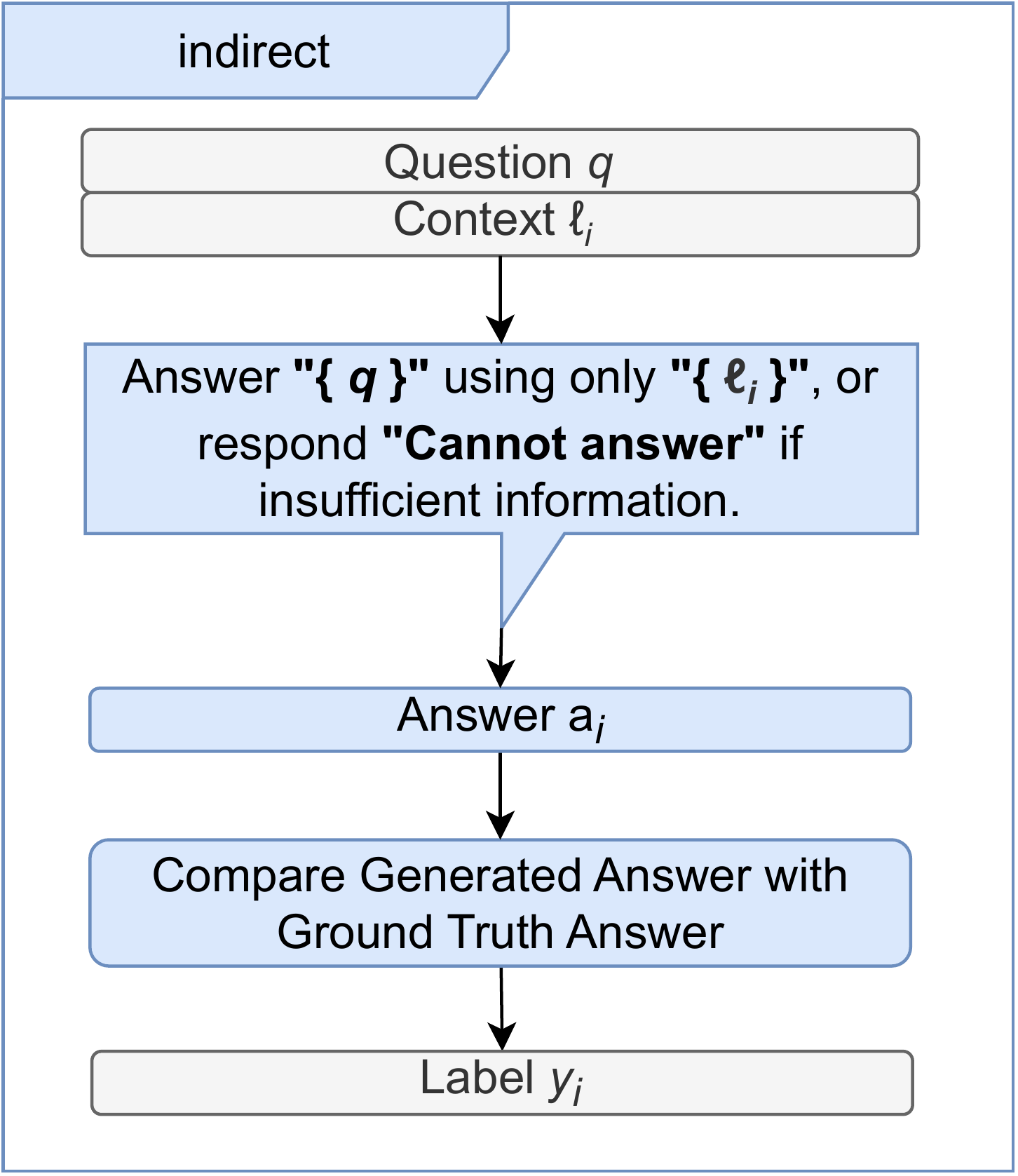}
    \caption{Indirect Approach}
    \label{fig:indirect-model}
  \end{subfigure}
  \hfill
  \begin{subfigure}[b]{0.3\textwidth}
    \includegraphics[width=\linewidth]{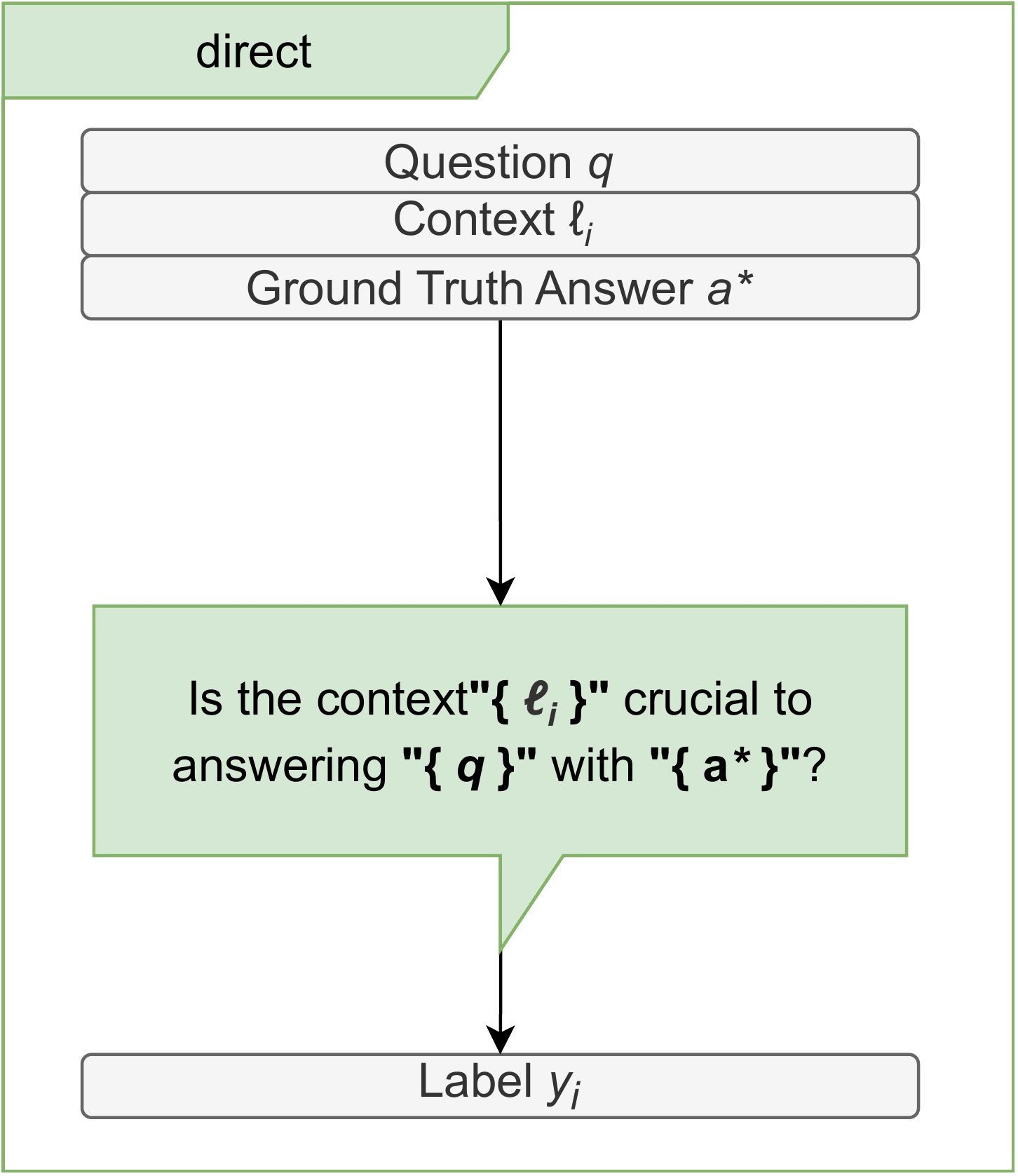}
    \caption{Direct Approach}
    \label{fig:direct-model}
  \end{subfigure}
  \hfill
  \begin{subfigure}[b]{0.3\textwidth}
    \includegraphics[width=\linewidth]{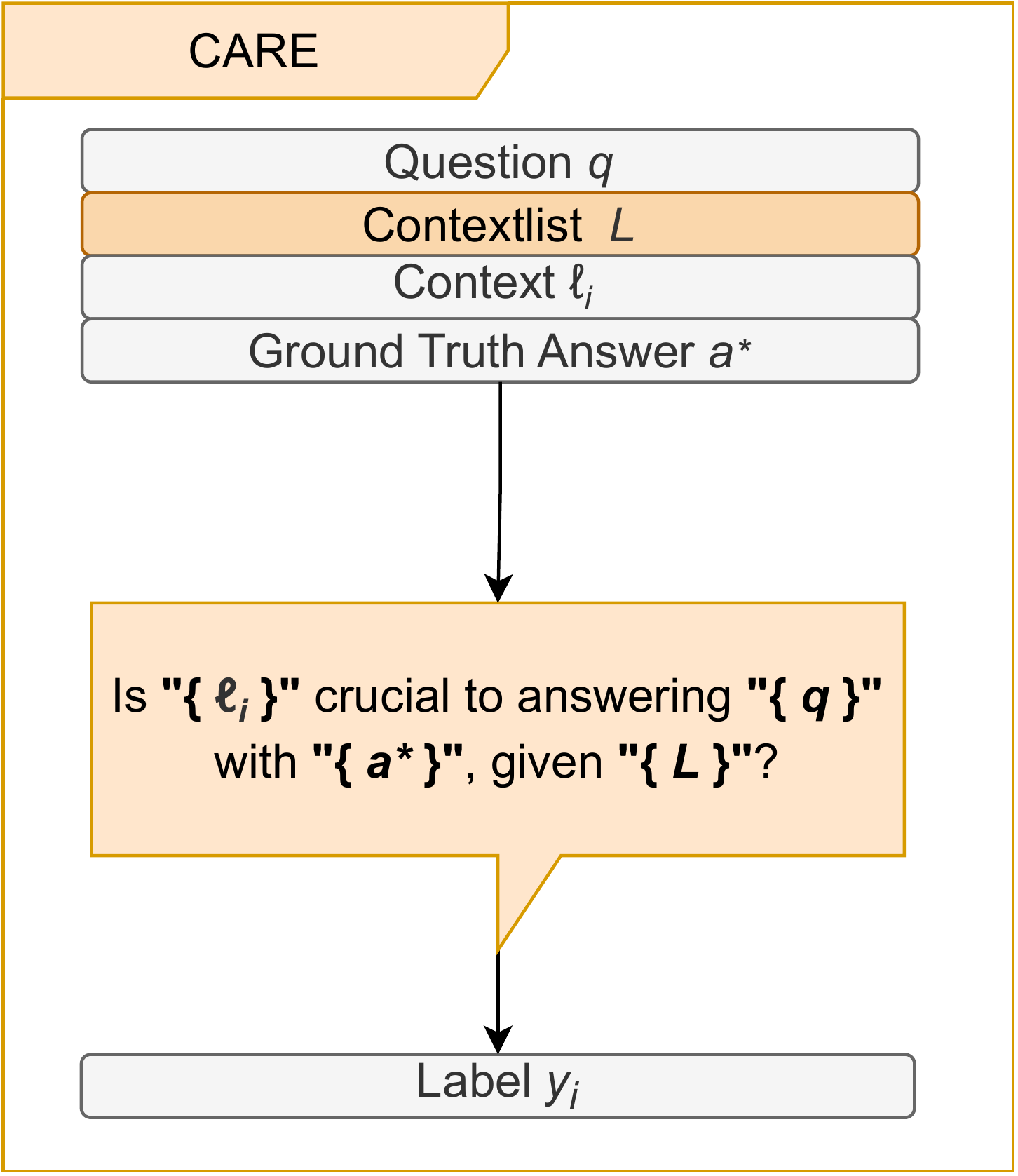}
    \caption{CARE Approach}
    \label{fig:care-model}
  \end{subfigure}
  
  \caption{Illustration of considered evaluation strategies}
  \label{fig:three-strategies}

\end{figure}

In this work, we compare three retrieval evaluation approaches:
\begin{enumerate*}[label=(\roman*)]
    \item an indirect evaluation approach derived from the \textit{eRAG} method~\cite{salemi_evaluating_2024}, and
    \item a direct evaluation method based on the ARES framework~\cite{saad-falcon_ares_2024},
    \item our proposed context-aware retriever evaluation (CARE) method 
\end{enumerate*}
(see Figure~\ref{fig:three-strategies}).

Given a query $q$, the retriever $R$ returns a list of context documents $\mathcal{L} = \{\ell_1, \ell_2, \dots, \ell_n\}$, where $\ell_i \in \mathcal{L}$ denotes a single retrieved context. Let $a^\ast$ denote the ground-truth answer associated with $q$. Each evaluation method outputs a binary relevance label $y_i \in \{0,1\}$, where $y_i = 1$ denotes relevance of context $\ell_i$ with respect to query $q$, and $y_i = 0$ denotes non-relevance.

In the \textit{indirect method}~\cite{salemi_evaluating_2024}, an LLM $\mathcal{I}$ attempts to answer the query $q$ using only a single context document $\ell_i$, producing an answer $a_i = \mathcal{I}(q, \ell_i)$. If $a_i = \varnothing$, the context $\ell_i$ is labeled as non-relevant. Otherwise, $a_i$ is compared to the ground-truth answer $a^\ast$ using an exact-match function, and the context is labeled as relevant if the answers match, and non-relevant otherwise. To reduce errors and biases introduced by strict exact matching, we manually reviewed all cases where no match was detected. Both the pre-review and revised labels are available in our repository. In practice, the comparison criterion may be adapted to the target application, ranging from strict exact matching to more flexible approaches such as LLM-based semantic equivalence judgments.

The \textit{direct method}~\cite{saad-falcon_ares_2024} uses an LLM $\mathcal{D}$ to predict the relevance label directly, defined as $y_i = \mathcal{D}(q, a^\ast, \ell_i)$. The LLM is provided with a detailed description of the task, including explicit instructions on how to assess the relevance of a context. The LLM receives several pieces of information as input: (1) the question being addressed $q$, (2) the specific context document $\ell_i$ to be classified, and (3) the ground-truth answer $a^\ast$. Based on this information, the LLM classifies the context document as either relevant or non-relevant.

Our proposed \textit{CARE} method extends the direct evaluation with a fourth element, (4) the list of context documents $\mathcal{L}$. Specifically, relevance is predicted as $y_i = \mathcal{C}(q, a^\ast, \mathcal{L}, \ell_i)$. The inclusion of the context list is particularly important in multi-hop settings, as it allows the evaluation to account for the presence of multiple supporting contexts.

The complete approaches are illustrated in Figure \ref{fig:three-strategies}, and the prompts used are available in our repository~\cite{noauthor_general_nodate}. We initially employed a zero-shot prompting strategy to ensure that the task was explained as clearly as possible. 
 These approaches assume that, in the RAG evaluation process, a dataset with questions and their corresponding ground-truth answers is available, and that a question may require multiple contexts to generate an answer. As a result, the evaluation output consists of a label indicating the relevance of each context.

\section{Experimental Setup}

\textit{Datasets and sampling.}
For our experiments, we used three datasets, the HotPotQA, SQuAD 2.0 and MuSiQue datasets. From each, we sampled 600 relevant and 600 non-relevant contexts. For HotPotQA, we stratified by difficulty (easy, medium, hard), selecting 400 contexts per level. Within each level, we balanced reasoning type with 200 comparison and 200 bridge contexts, yielding six categories of 200 samples each.

\textit{Models.} For SQuAD 2.0 and MuSiQue, we performed experiments exclusively with GPT-4.1. To assess differences in model performance, we conducted evaluations using six LLMs on the HotPotQA dataset, from Gemini, OpenAI, and Meta’s LLaMa series using their standard settings. From Gemini, we used the gemini-2.0-flash as the standard model and the gemini-1.5-flash-8b as the smaller model via Google’s API~\cite{noauthor_gemini_nodate}. From Meta, we included the LLaMa 3.1–70b as a medium-sized model and the LLaMa 3.1–8b as the smaller model ~\cite{LLaMa3modelcard}. From OpenAI, we selected the standard GPT-4.1 model, as well as the o4-mini, which we considered as our reasoning model through the OpenAI API~\cite{noauthor_model_nodate}.
This setup enabled us to compare model sizes, reasoning capabilities, and the viability of open-source LLaMa models for self-hosted use.

\textit{Prompting.}
Our primary objective was to compare evaluation methods through zero-shot prompting rather than fine-tuning models. Labels were generated using the prompt templates provided in the repository, and we deliberately avoided model-specific prompt optimization to ensure independence and generalizability. However, for the CARE evaluation, we also examined the impact of different prompting strategies on HotPotQA by modifying the original prompt and testing various configurations.

\textit{Metrics and statistical analysis.} We computed standard retrieval metrics including accuracy, precision, recall, and F1-Score to assess performance. To enhance the robustness of our findings, we applied bootstrapping techniques~\cite{efron_introduction_1994}. When bootstrapping the entire dataset, we generated 5,000 samples, each containing 1,200 data points. For bootstrapping specific question types or difficulty levels, we used the corresponding subset sizes (e.g., 600 for bridge questions, 400 for the easy difficulty level). From these samples, we constructed 95\% confidence intervals using an $\alpha$ level of 0.05. To assess the statistical significance and robustness of our results, we conducted permutation tests with 9{,}999 permutations.

\section{Results}

In the following section, we present our experimental results.
All tables report bootstrapped confidence intervals and indicate statistically significant results from permutation tests with an asterisk (*) relative to the highlighted baseline (Base). For the CARE approach, we indicate the number of contexts in the context list by appending it as a subscript in the format CARE\textsubscript{n}, where $n$ denotes the size of the list.

\textit{Multi-Hop.}
Our initial results demonstrate in Table \ref{tab:model_comparison_gpt4.1} the robustness of CARE for multi-hop queries. It achieves significantly better performance compared to both the indirect and direct approaches across nearly all metrics, including accuracy, F1-Score, and recall, in both multi-hop datasets. Both the indirect and direct approaches show higher precision than recall, indicating that when they classify a context as relevant, it is likely to be correct. However, their relatively low recall suggests they miss a significant portion of relevant contexts, likely labeling them as relevant only when highly confident. Overall, the direct approach outperforms the indirect approach, achieving significantly higher accuracy, F1-Score, and recall.

\begin{table}[ht]
    \centering
    \caption{Comparison of different approaches for GPT-4.1}
    \begin{tabular}{p{2cm}p{2cm}p{2cm}p{2cm}p{2cm}}
        \toprule
        \textbf{Approach} & \textbf{Accuracy} & \textbf{F1-Score} & \textbf{Recall} & \textbf{Precision} \\
        \midrule
         \multicolumn{5}{c}{\textbf{HotPotQA}} \\
           \midrule 
        Indirect\cite{salemi_evaluating_2024}  & \textbf{0.642$\pm$0.03*} & \textbf{0.474$\pm$0.04*} & \textbf{0.322$\pm$0.04*} & 0.898$\pm$0.04 \\
        Direct\cite{saad-falcon_ares_2024}  & \textbf{0.720$\pm$0.03*} & \textbf{0.658$\pm$0.04*} & \textbf{0.540$\pm$0.04*} & 0.844$\pm$0.04 \\
        CARE\textsubscript{10}\textsuperscript{Base}   & 0.827$\pm$0.02& 0.814$\pm$0.02 & 0.757$\pm$0.04 & 0.880$\pm$0.03 \\
        \midrule
         \multicolumn{5}{c}{\textbf{MuSiQue}} \\
           \midrule 
        Indirect\cite{salemi_evaluating_2024}  & \textbf{0.631$\pm$0.03*} & \textbf{0.417$\pm$0.04*} & \textbf{0.264$\pm$0.03*} & 0.994$\pm$0.01 \\
        Direct\cite{saad-falcon_ares_2024}  & \textbf{0.702$\pm$0.03*} & \textbf{0.578$\pm$0.04*} & \textbf{0.410$\pm$0.04*} & 0.984$\pm$0.01 \\
        CARE\textsubscript{20}\textsuperscript{Base}   & 0.755$\pm$0.02& 0.678$\pm$0.03 & 0.517$\pm$0.04 & 0.987$\pm$0.01 \\
        \bottomrule
    \end{tabular}

 \caption*{\footnotesize * statistically significant difference; \textsuperscript{Base} denotes baseline.}
    \label{tab:model_comparison_gpt4.1}
\end{table}

\textit{LLM model comparison.} The experiments were replicated on the HotPotQA dataset using various underlying models for comparison as presented in table \ref{tab:model_comparison}. 
The indirect approach led to a significant improvement in F1-Score for the small LLaMa model. In contrast, the direct approach resulted in a decline in F1-Score for the reasoning model o4-mini. However, both approaches generally performed comparably across different models.
For CARE, the reasoning model o4-mini exhibited a decrease in accuracy, F1-Score, and recall compared to GPT-4.1, while precision significantly improved. Among smaller models, we observed a shift from precision to recall. Notably, the LLaMa 3.1-8b model experienced a significant decline in overall performance, with substantial drops in both F1-Score and accuracy. Unexpectedly, its accuracy fell below that of the direct approach. Furthermore, CARE consistently outperformed other approaches across all models except for the LLaMa 3.1-8b model, demonstrating stable performance when context size and parameter size were sufficiently high.

 \begin{table}[ht]
    \centering
    \caption{Model comparison on the HotPotQA dataset.} 
    \begin{tabular}{p{1.9cm}p{2cm}p{1.9cm}p{1.9cm}p{1.9cm}p{1.9cm}}
        \toprule
        \textbf{Approach} & \textbf{Model} &  \textbf{Accuracy} & \textbf{F1-Score} & \textbf{Recall} & \textbf{Precision} \\ 
          \toprule 
         \textbf{Indirect}~\cite{salemi_evaluating_2024}&GPT-4.1\textsuperscript{Base}  & 0.642$\pm$0.03 & 0.474$\pm$0.04 & 0.322$\pm$0.04 & 0.898$\pm$0.04 \\
        & o4-mini   & 0.643$\pm$0.03 & 0.473$\pm$0.04 & 0.322$\pm$0.04 & 0.898$\pm$0.04 \\
         \midrule 
         &2.0-flash\textsuperscript{Base}       & 0.595$\pm$0.03 & 0.345$\pm$0.05 & 0.213$\pm$0.03 & 0.901$\pm$0.05 \\
        & 1.5-flash-8b    & 0.579$\pm$0.03 & 0.294$\pm$0.04 & 0.175$\pm$0.03 & 0.914$\pm$0.05 \\
        \midrule
        & LLaMa70b\textsuperscript{Base}  & 0.605$\pm$0.03 & 0.368$\pm$0.05 & 0.230$\pm$0.03 & 0.920$\pm$0.05 \\
        & LLaMa8b & 0.566$\pm$0.03 & \textbf{0.445$\pm$0.04*} & \textbf{0.348$\pm$0.04*} & \textbf{0.618$\pm$0.05*} \\
          
          \toprule 
        
         \textbf{Direct}~\cite{saad-falcon_ares_2024} & GPT-4.1\textsuperscript{Base}
        & 0.720$\pm$0.03 & 0.658$\pm$0.04 & 0.540$\pm$0.04 & 0.844$\pm$0.04 \\
        & o4-mini
        & 0.694$\pm$0.03 & \textbf{0.590$\pm$0.04*} & \textbf{0.442$\pm$0.04*} & 0.889$\pm$0.04 \\      
        \midrule 
        & 2.0-flash\textsuperscript{Base}
        &0.691$\pm$0.03 &0.590$\pm$0.04& 0.443$\pm$0.04 & 0.881$\pm$0.04 \\
        & 1.5-flash-8b 
        & 0.708$\pm$0.03 & 0.610$\pm$0.04 & 0.457$\pm$0.04 & 0.920$\pm$0.03 \\
        \midrule 
           & LLaMa70b\textsuperscript{Base}
          &0.745$\pm$0.03 & 0.686$\pm$0.03 & 0.556$\pm$0.04 & 0.895$\pm$0.03 \\
       & LLaMa8b
          & 0.748$\pm$0.03 & 0.719$\pm$0.03 & \textbf{0.643$\pm$0.04*} & \textbf{0.815$\pm$0.04*} \\
      
          \toprule
          \textbf{CARE\textsubscript{10}} &  GPT-4.1\textsuperscript{Base}
           & 0.827$\pm$0.02 & 0.814$\pm$0.02 & 0.757$\pm$0.04 & 0.880$\pm$0.03 \\
        & o4-mini
         & \textbf{0.781$\pm$0.02*} & \textbf{0.732$\pm$0.03*} & \textbf{0.600$\pm$0.04*} & \textbf{0.940$\pm$0.03*} \\  
         \midrule
         & 2.0-flash\textsuperscript{Base}
             & 0.840$\pm$0.02 & 0.825$\pm$0.04 &0.753$\pm$0.04 & 0.911$\pm$0.03 \\
        & 1.5-flash-8b 
         & 0.812$\pm$0.02 & 0.822$\pm$0.02 & \textbf{0.872$\pm$0.03*} & \textbf{0.778$\pm$0.03*} \\
         \midrule 
       & LLaMa70b\textsuperscript{Base}
          & 0.833$\pm$0.02 &0.815$\pm$0.03 & 0.738$\pm$0.03 & 0.910$\pm$0.03 \\
       & LLaMa8b
       & \textbf{0.674$\pm$0.03*} & \textbf{0.736$\pm$0.03*} & \textbf{0.908$\pm$0.02*} & \textbf{0.619$\pm$0.03*} \\
       \bottomrule
    \end{tabular}
    \label{tab:model_comparison}
 \caption*{\footnotesize * statistically significant difference; \textsuperscript{Base} denotes baseline.}
\end{table}

\textit{Question Levels, Types, and Overall Performance.} Table \ref{tab:type_level_comparison} presents the comparison of F1-Scores using GPT-4.1 across all difficulty levels (easy, medium, hard), question types (bridge and comparison). The results show that, regardless of question type or difficulty level, CARE consistently achieved the highest performance across all subsets and outperformed other approaches on the entire dataset. Additionally, the direct approach exhibits a performance drop on medium and hard questions, while both direct and indirect approaches perform worse on comparison questions. The indirect approach does not show a significant performance drop on medium and hard questions, but generally performs worse across all difficulty levels, including easy ones.

\begin{table}[ht]
    \centering
    \caption{F1-Score Comparison: HotPotQA Levels and Question (GPT-4.1)}
    \begin{tabular}{p{2cm}p{2cm}p{2cm}p{2cm}}
    \toprule
        \textbf{Category} & \textbf{Indirect~\cite{salemi_evaluating_2024}} & \textbf{Direct~\cite{saad-falcon_ares_2024}} & \textbf{CARE}  \\
        \midrule
        
          Easy\textsuperscript{Base} & 0.530 $\pm$ 0.07&0.758 $\pm$ 0.05 & 0.808 $\pm$ 0.04\\
          Medium & 0.444$\pm$0.08 & \textbf{0.637$\pm$0.06*} &0.822 $\pm$ 0.04\\
          Hard & 0.438$\pm$0.08 & \textbf{0.556$\pm$0.07*} & 0.809 $\pm$ 0.05 \\
         
        \midrule

        Bridge\textsuperscript{Base} & 0.639$\pm$0.05 &  0.724$\pm$0.05 & 0.804$\pm$0.04\\
        Comparison  & \textbf{0.264$\pm$0.06*} & \textbf{0.591$\pm$0.05*} & 0.822$\pm$0.03 \\
        \midrule
        
    \end{tabular}
    \label{tab:type_level_comparison}
     \caption*{\footnotesize * statistically significant difference; \textsuperscript{Base} denotes baseline.}
\end{table}

\textit{Prompting strategies.}
Furthermore, we explored various prompting strategies
through additional experiments using the CARE approach. First, we applied single-shot prompting, in which only one example is included in the prompt to guide the model’s reasoning~\cite{brown_language_2020}. We then extended this to few-shot prompting by providing at least two examples--one with a relevant context and one with a non-relevant context~\cite{brown_language_2020}. As a third strategy, we employed a role-based prompt, assigning the LLM the role of an evaluator and explicitly instructing it on how to assess relevance~\cite{wu_large_2023}. Another variant involved minimal instruction, presenting only the essential input information without any contextual guidance. Finally, we experimented with the Chain-of-Thought (CoT) technique~\cite{wei_chain--thought_2023}, combining it with few-shot prompting to encourage step-by-step reasoning (cf. repository for all prompting templates).
To analyze the prompting strategies, we compared the alternative prompting strategies against the standard prompt used in our previous experiments (cf. Table \ref{tab:prompting_comparison}). The short prompt yielded a significant improvement across all metrics except precision. The role-based prompt also led to a significant increase in F1-Score and recall. Only the CoT prompt resulted in a significant decline in all metrics except precision. This shows similarity to the reasoning model, which uses CoT techniques in the background, o4-mini. The remaining prompting strategies did not produce any statistically significant changes in performance.

\begin{table}[ht]

\centering
\begin{minipage}[t]{0.43\textwidth}
    \centering
    \caption{\footnotesize{CARE Prompting Strategies on HotPotQA (GPT-4.1)}}
    \begin{tabular}{p{2cm}p{2.5cm}}
        \toprule
        \textbf{Prompt} & \textbf{F1-Score} \\
        \midrule
        standard\textsuperscript{Base}   & 0.814$\pm$0.02  \\
        \midrule
        single-shot    & 0.818$\pm$0.02 \\
        few-shot     & 0.798$\pm$0.03  \\
        role     & \textbf{0.849$\pm$0.02*}  \\
        short     & \textbf{0.877$\pm$0.02*}   \\
        CoT     &  \textbf{0.722$\pm$0.03*}  \\
        \bottomrule
    \end{tabular}
    \label{tab:prompting_comparison}
\end{minipage}
\hfill
\begin{minipage}[t]{0.43\textwidth}
    \centering
    \caption{\footnotesize{CARE and Direct (with/without answer) on HotPotQA using Gemini-2.0-flash}}
    \begin{tabular}{p{2cm}p{2.5cm}}
        \toprule
        \textbf{Approach} &  \textbf{F1-Score}\\
        \midrule
       
        Direct\cite{saad-falcon_ares_2024}\textsuperscript{Base}      & 0.590$\pm$0.04 \\
        Direct$^\dagger$\cite{saad-falcon_ares_2024} &   \textbf{0.488$\pm$0.04*}  \\
        \midrule
        CARE\textsubscript{10}\textsuperscript{Base}     & 0.825$\pm$0.04\\
        CARE\textsubscript{10}$^\dagger$     & 0.869$\pm$0.02   \\
        \bottomrule
    \end{tabular}
    
    \label{tab:answer_comparison}
\end{minipage}
\vspace{1em}
 \caption*{\footnotesize   * statistically significant difference;
 \textsuperscript{Base} denotes baseline; $^\dagger$ Approach without answer.}

\end{table}

To better understand the poor performance of the CoT prompting approach, we manually analyzed the reasoning steps generated by the model. 
Our manual analysis of the CoT prompt errors identified three main types: hallucination errors (40 instances), where the model invented information; context awareness errors (214 instances), where it failed to use the provided context effectively; and context interpretation errors (26 instances), where the model's relevance judgment differed from the ground-truth.


\textit{Without answers.}
To account for scenarios in which the evaluation dataset lacks a reference answer, we adapted both CARE and the Direct approach by removing the ground-truth answer from the prompt. We then evaluated the performance of both methods, the Direct approach and CARE, under this setting, compare Table \ref{tab:answer_comparison}. While excluding the ground-truth answer led to a decline in all metrics except precision for the Direct approach, CARE showed no significant negative impact.

\begin{figure}[ht]
    \begin{minipage}{0.38\textwidth}
    \centering

    \centering
      \captionof{table}{\footnotesize{CARE and Direct on SQuAD2.0 (GPT-4.1)}} 
    \begin{tabular}{p{2cm}p{2cm}}
        \toprule
        \textbf{Approach} &  \textbf{F1-Score}\\ 
          \midrule

         Direct\cite{saad-falcon_ares_2024}& \textbf{0.819$\pm$0.02*} \\
         \midrule 
         CARE\textsubscript{0}   &\textbf{0.795$\pm$0.02*}  \\
         CARE\textsubscript{1}     & 0.748$\pm$0.02 \\
        CARE\textsubscript{3}   & 0.739$\pm$0.02  \\
        CARE\textsubscript{5}    & 0.738$\pm$0.02  \\
        CARE\textsubscript{10}\textsuperscript{Base}   & 0.735$\pm$0.02  \\
        CARE\textsubscript{20}  & 0.751$\pm$0.02 \\
       \bottomrule
       
    \end{tabular}
    \label{tab:sinlgehop_comparison}
    \end{minipage}
    \hfill
    \begin{minipage}{0.58\textwidth}
        \centering
        \includegraphics[width=\linewidth]{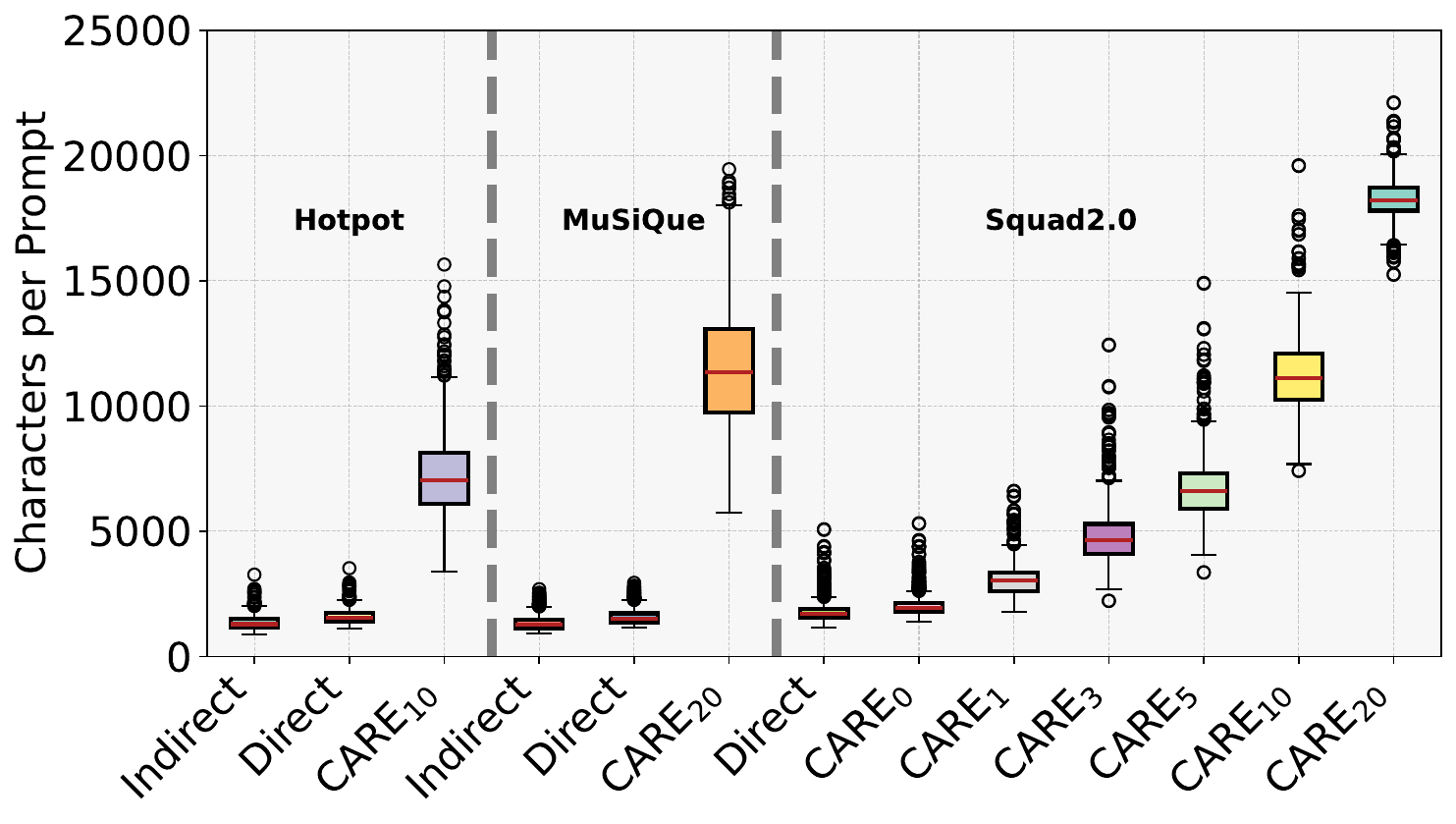}
        \caption{\footnotesize{Prompt length by approach}}
        \label{fig:avg_chars}
    \end{minipage}
     \caption*{\footnotesize * statistically significant difference; \textsuperscript{Base} denotes baseline.}
    \end{figure}

\textit{Single-Hop.}
To assess whether the CARE approach--despite its longer prompts and higher computational cost, is still suitable for single-hop queries, we compared it to the direct evaluation method. Due to the lack of ground-truth answers for unanswerable questions in the SQuAD 2.0 dataset, we excluded the indirect approach from this comparison and used the slightly modified approach where the answer is not included like in \ref{tab:answer_comparison}. In this setup, we observed a significant drop in CARE’s performance.

\textit{Contextlist length.}
To gain deeper insight into the CARE approach, we also conducted an ablation study to examine the impact of context list length on performance. Specifically, we evaluated CARE using context lists containing 0, 1, 3, 5, 10, and 20 documents (cf. Table \ref{tab:sinlgehop_comparison}). Notably, only the configuration with zero context documents--closely resembling the direct evaluation setup--achieved significantly better results. For all other configurations, performance did not differ significantly, indicating that the number of context documents in the list has limited impact on CARE's effectiveness in this setting.
These results are summarized in Table \ref{tab:sinlgehop_comparison}.

\textit{Prompt length.}
We compared the number of characters per prompt and found that CARE prompts were over four times longer than those of other methods (cf. Figure \ref{fig:avg_chars}). The length depends on the size and number of retrieved contexts included in the prompt. 

\section{Discussion}

We evaluated three LLM-as-judge retriever evaluation strategies in a synthetic RAG setup on HotPotQA, MuSiQue, and SQuAD 2.0, focusing on their ability to label context relevance, particularly for multi-hop reasoning in RAG systems.
Deciding on an appropriate evaluation method for a RAG system largely depends on the complexity of the target questions. Our findings indicate that for multi-hop queries, evaluators must consider the full set of retrieved contexts rather than assessing each context in isolation. Due to the nature of multi-hop reasoning, individual contexts may appear irrelevant on their own, while their combination renders them jointly relevant to the query. CARE leverages this interdependence by evaluating contexts in light of the complete context list, which explains its improved performance.
By contrast, for single-hop questions, relevance is typically self-contained within a single context, which can be assessed independently without requiring additional supporting contexts. In this setting, the direct approach is generally more effective: it achieves higher performance than CARE while being less computationally expensive due to its simpler structure and shorter prompt length. The indirect approach is mainly effective on the easy subset of HotPotQA, which is largely single-hop. In this setting, both direct and indirect methods outperform their results on more complex multi-hop questions, consistent with \cite{salemi_evaluating_2024}, who report strong indirect performance under similar conditions. However, its effectiveness drops sharply on multi-hop questions, whereas CARE maintains stable performance across both single-hop and multi-hop queries, even when handling longer context lists. This robustness makes CARE a broadly applicable and reliable evaluation method.

When evaluating a RAG system, it is important to consider the structure of the evaluation dataset. This includes whether contexts are explicitly linked to questions and ground-truth answers, and whether true answers are available at all. CARE is a robust and practical solution for both labeled and unlabeled datasets. By default, it is designed for unlabeled datasets, as it can generate its own relevance labels. However, it is also valuable for labeled datasets, where it addresses a common limitation: the lack of explicit ``not relevant'' labels. As noted by Salemi et al.~\cite{salemi_evaluating_2024}, a context can be relevant even if no label is assigned, while irrelevant contexts are often unannotated~\cite{tang_multihop-rag_2024,es_ragas_2023}. Moreover, our experiments demonstrate that the absence of answers does not reduce its performance. This makes CARE a practical solution for automating retriever evaluation without relying on large labeled datasets or ground-truth answers, while also enabling integration into live systems for real-time, continuous relevance assessment.
A surprising finding was that reasoning models reduced CARE's evaluation performance, a result further confirmed by the CoT prompting experiment. These results suggest that, at present, reasoning-heavy approaches are ill-suited for retriever evaluation with CARE. Our qualitative analysis indicates that such models often hallucinate, lose track of the task, or fail to consider the full context. While future LLM improvements may mitigate these issues, our experiments show that current reasoning models and CoT prompting do not perform well within the CARE framework.

A key limitation of CARE is its incompatibility with smaller LLMs in terms of parameters and context length, whereas the direct and indirect approaches remain usable. This is evident with o4-mini and LLaMa 3.1–8B, where even direct evaluation outperforms CARE. These models, having fewer parameters and shorter context windows, struggle with CARE’s reliance on complex prompts and extended input. Since effective use of long contexts requires strong comprehension and retention, these findings align with prior work showing that not all models handle long-context reasoning effectively~\cite{bai_longbench_2025,li_long-context_2024}. An advantage of smaller open-source LLaMa models is that they can be self-hosted on modest hardware, making them suitable for cost-sensitive deployments. This raises the question of whether CARE requires the full capabilities of high-performance models, or if smaller models can offer a cost- and energy-efficient alternative. Our findings suggest that this depends on the specific model: for instance, LLaMa 3.1–8B underperformed compared to the medium-sized variant, yet smaller Gemini models delivered acceptable results. This indicates that CARE can, under certain conditions, be executed effectively on less powerful models, reducing both cost and energy consumption. By contrast, model size and quality had only minimal impact on the direct and indirect approaches. While CARE places slightly higher demands due to complex prompts and extended context, smaller models can still meet these requirements with the right configuration. 

\section{Conclusion}
This paper demonstrates that context-aware evaluation is essential for RAG systems, particularly for multi-hop queries where assessing the full retrieved context set is critical. Our findings reveal that competitive evaluation performance does not strictly necessitate the largest models. Instead, it depends on sufficient context windows and reasoning capabilities. This indicates that cost-efficient models can be effective evaluators if they meet these baseline requirements.
Ultimately, the choice of evaluation strategy involves a trade-off: while our proposed method (CARE) offers superior reliability for complex tasks, simpler direct or indirect methods may remain more efficient for single-hop queries or resource-constrained environments.
Our results are primarily constrained by their reliance on synthetic, text-only datasets. In production environments, RAG systems frequently encounter unstructured data and multimodal inputs, such as images and tables \cite{brehme_retrieval-augmented_2025}. Therefore, our findings are specific to well-processed textual contexts. Future research should investigate the efficacy of CARE as a live evaluator in real-world settings, particularly its robustness on uncleaned and multimodal data. Moreover, a more systematic ablation and hyperparameter study is necessary to better understand why CARE performs better. This should include analyses of the number and difficulty of support contexts as well as the contribution of multi-hop reasoning, especially given its weaker performance in single-hop cases and when used with reasoning-focused models.

\subsubsection*{Acknowledgements.} The authors would like to thank Steffen Hahn, Andreas Egger, Mateo Golec, and Daniel Walder for their initial efforts related to this work.


\bibliographystyle{splncs04}
\bibliography{references}

@misc{rackauckas_evaluating_2024,
	title = {Evaluating {RAG}-Fusion with {RAGElo}: an Automated Elo-based Framework},
	url = {http://arxiv.org/abs/2406.14783},
	doi = {10.48550/arXiv.2406.14783},
	shorttitle = {Evaluating {RAG}-Fusion with {RAGElo}},
	number = {{arXiv}:2406.14783},
	publisher = {{arXiv}},
	author = {Rackauckas, Zackary and Câmara, Arthur and Zavrel, Jakub},
	urldate = {2024-11-06},
	date = {2024-10-08},
    year ={2024},
}

@misc{friel_ragbench_2024,
	title = {{RAGBench}: Explainable Benchmark for Retrieval-Augmented Generation Systems},
	url = {http://arxiv.org/abs/2407.11005},
	doi = {10.48550/arXiv.2407.11005},
	shorttitle = {{RAGBench}},
	number = {{arXiv}:2407.11005},
	publisher = {{arXiv}},
	author = {Friel, Robert and Belyi, Masha and Sanyal, Atindriyo},
	urldate = {2024-11-06},
	date = {2024-06-25},
    year ={2024},
}

@misc{ding_vera_2024,
	title = {{VERA}: Validation and Evaluation of Retrieval-Augmented Systems},
	url = {http://arxiv.org/abs/2409.03759},
	doi = {10.48550/arXiv.2409.03759},
	shorttitle = {{VERA}},
	number = {{arXiv}:2409.03759},
	publisher = {{arXiv}},
	author = {Ding, Tianyu and Banerjee, Adi and Mombaerts, Laurent and Li, Yunhong and Borogovac, Tarik and Weinstein, Juan Pablo De la Cruz},
	urldate = {2024-12-09},
	date = {2024-08-16},
    year = {2024},
	eprinttype = {arxiv},
}

@misc{saad-falcon_ares_2024,
	title = {{ARES}: An Automated Evaluation Framework for Retrieval-Augmented Generation Systems},
	url = {http://arxiv.org/abs/2311.09476},
	doi = {10.48550/arXiv.2311.09476},
	shorttitle = {{ARES}},
	number = {{arXiv}:2311.09476},
	publisher = {{arXiv}},
	author = {Saad-Falcon, Jon and Khattab, Omar and Potts, Christopher and Zaharia, Matei},
	urldate = {2024-11-06},
	date = {2024-03-31},
    year ={2024},
}

@misc{salemi_evaluating_2024,
	title = {Evaluating Retrieval Quality in Retrieval-Augmented Generation},
	url = {http://arxiv.org/abs/2404.13781},
	doi = {10.48550/arXiv.2404.13781},
	number = {{arXiv}:2404.13781},
	publisher = {{arXiv}},
	author = {Salemi, Alireza and Zamani, Hamed},
	urldate = {2024-11-06},
	date = {2024-04-21},
    year ={2024},
}

@misc{alinejad_evaluating_2024,
	title = {Evaluating the Retrieval Component in {LLM}-Based Question Answering Systems},
	url = {http://arxiv.org/abs/2406.06458},
	doi = {10.48550/arXiv.2406.06458},
	number = {{arXiv}:2406.06458},
	publisher = {{arXiv}},
	author = {Alinejad, Ashkan and Kumar, Krtin and Vahdat, Ali},
	urldate = {2024-12-09},
	date = {2024-06-10},
    year ={2024},
	eprinttype = {arxiv},
}

@misc{es_ragas_2023,
	title = {{RAGAS}: Automated Evaluation of Retrieval Augmented Generation},
	url = {http://arxiv.org/abs/2309.15217},
	doi = {10.48550/arXiv.2309.15217},
	shorttitle = {{RAGAS}},
	number = {{arXiv}:2309.15217},
	publisher = {{arXiv}},
	author = {Es, Shahul and James, Jithin and Espinosa-Anke, Luis and Schockaert, Steven},
	urldate = {2024-11-06},
	date = {2023-09-26},
    year ={2023},
	eprinttype = {arxiv},
}

@misc{gao_retrieval-augmented_2024,
	title = {Retrieval-Augmented Generation for Large Language Models: A Survey},
	url = {http://arxiv.org/abs/2312.10997},
	doi = {10.48550/arXiv.2312.10997},
	shorttitle = {Retrieval-Augmented Generation for Large Language Models},
	number = {{arXiv}:2312.10997},
	publisher = {{arXiv}},
	author = {Gao, Yunfan and Xiong, Yun and Gao, Xinyu and Jia, Kangxiang and Pan, Jinliu and Bi, Yuxi and Dai, Yi and Sun, Jiawei and Wang, Meng and Wang, Haofen},
	urldate = {2025-01-14},
	date = {2024-03-27},
    year ={2024},
	eprinttype = {arxiv},
}

@misc{lewis_retrieval-augmented_2021,
	title = {Retrieval-{Augmented} {Generation} for {Knowledge}-{Intensive} {NLP} {Tasks}},
	url = {http://arxiv.org/abs/2005.11401},
	urldate = {2024-10-14},
	publisher = {arXiv},
	author = {Lewis, Patrick and Perez, Ethan and Piktus, Aleksandra and Petroni, Fabio and Karpukhin, Vladimir and Goyal, Naman and Küttler, Heinrich and Lewis, Mike and Yih, Wen-tau and Rocktäschel, Tim and Riedel, Sebastian and Kiela, Douwe},
	month = apr,
	year = {2021},
	note = {arXiv:2005.11401},
}

@misc{tang_multihop-rag_2024,
	title = {{MultiHop}-{RAG}: Benchmarking Retrieval-Augmented Generation for Multi-Hop Queries},
	url = {http://arxiv.org/abs/2401.15391},
	doi = {10.48550/arXiv.2401.15391},
	shorttitle = {{MultiHop}-{RAG}},
	number = {{arXiv}:2401.15391},
	publisher = {{arXiv}},
	author = {Tang, Yixuan and Yang, Yi},
	urldate = {2024-11-06},
	date = {2024-01-27},
    year ={2024},
}

@misc{afzal_towards_2024,
	title = {Towards Optimizing and Evaluating a Retrieval Augmented {QA} Chatbot using {LLMs} with Human in the Loop},
	url = {http://arxiv.org/abs/2407.05925},
	doi = {10.48550/arXiv.2407.05925},
	number = {{arXiv}:2407.05925},
	publisher = {{arXiv}},
	author = {Afzal, Anum and Kowsik, Alexander and Fani, Rajna and Matthes, Florian},
	urldate = {2024-12-09},
	date = {2024-07-08},
    year = {2024},
	eprinttype = {arxiv},
}

@misc{yang2018hotpotqa,
  title = {{HotpotQA}: A Dataset for Diverse, Explainable Multi-hop Question Answering},
	url = {http://arxiv.org/abs/1809.09600},
	doi = {10.48550/arXiv.1809.09600},
	shorttitle = {{HotpotQA}},
	number = {{arXiv}:1809.09600},
	publisher = {{arXiv}},
	author = {Yang, Zhilin and Qi, Peng and Zhang, Saizheng and Bengio, Yoshua and Cohen, William W. and Salakhutdinov, Ruslan and Manning, Christopher D.},
	urldate = {2025-05-06},
	date = {2018-09-25},
    year = {2018},
	eprinttype = {arxiv},
}

@misc{liu_cofe-rag_2024,
	title = {{CoFE}-{RAG}: A Comprehensive Full-chain Evaluation Framework for Retrieval-Augmented Generation with Enhanced Data Diversity},
	url = {http://arxiv.org/abs/2410.12248},
	doi = {10.48550/arXiv.2410.12248},
	shorttitle = {{CoFE}-{RAG}},
	number = {{arXiv}:2410.12248},
	publisher = {{arXiv}},
	author = {Liu, Jintao and Ding, Ruixue and Zhang, Linhao and Xie, Pengjun and Huang, Fie},
	urldate = {2024-11-06},
	date = {2024-10-16},
    year ={2024},
}

@misc{rajpurkar_know_2018,
	title = {Know What You Don't Know: Unanswerable Questions for {SQuAD}},
	url = {http://arxiv.org/abs/1806.03822},
	doi = {10.48550/arXiv.1806.03822},
	shorttitle = {Know What You Don't Know},
	number = {{arXiv}:1806.03822},
	publisher = {{arXiv}},
	author = {Rajpurkar, Pranav and Jia, Robin and Liang, Percy},
	urldate = {2025-01-16},
	date = {2018-06-11},
    year ={2024},
	eprinttype = {arxiv},
}

@misc{brehme2025SLR,
	title = {Can {LLMs} be Trusted for Evaluating {RAG} Systems? A Survey of Methods and Datasets},
	url = {https://ieeexplore.ieee.org/document/11081490},
	doi = {10.1109/SDS66131.2025.00010},
	shorttitle = {Can {LLMs} be Trusted for Evaluating {RAG} Systems?},
	eventtitle = {2025 {IEEE} Swiss Conference on Data Science ({SDS})},
	pages = {16--23},
	booktitle = {2025 {IEEE} Swiss Conference on Data Science ({SDS})},
	author = {Brehme, Lorenz and Ströhle, Thomas and Breu, Ruth},
	urldate = {2025-08-04},
	date = {2025-06},
        year = {2025},
	note = {{ISSN}: 2835-3420},
}

@misc{kukreja_performance_2024,
	title = {Performance Evaluation of Vector Embeddings with Retrieval-Augmented Generation},
	url = {https://ieeexplore.ieee.org/document/10603291},
	doi = {10.1109/ICCCS61882.2024.10603291},
	eventtitle = {2024 9th International Conference on Computer and Communication Systems ({ICCCS})},

	booktitle = {2024 9th International Conference on Computer and Communication Systems ({ICCCS})},
	author = {Kukreja, Sanjay and Kumar, Tarun and Bharate, Vishal and Purohit, Amit and Dasgupta, Abhijit and Guha, Debashis},
	urldate = {2024-10-25},
	date = {2024-04},
    year = {2024},
}

@misc{noauthor_model_nodate,
	title = {Model - {OpenAI} {API}},
	url = {https://platform.openai.com},
	urldate = {2025-04-28},
	langid = {english},
    author ={OpenAI},
    year = {2025},
}

@misc{noauthor_gemini_nodate,
	title = {Gemini models {\textbar} Gemini {API}},
	url = {https://ai.google.dev/gemini-api/docs/models},
	titleaddon = {Google {AI} for Developers},
	urldate = {2025-05-07},
	langid = {english},
    author = {Google DeepMind},
    year = {2025}
}

@misc{llama3modelcard,
  title={Llama 3.1 Model Card},
  author={AI@Meta},
  year={2024},
  url = {https://github.com/meta-llama/llama-models/blob/main/models/llama3\_1/MODEL\_CARD.md}
}

@book{efron_introduction_1994,
	location = {New York},
	title = {An Introduction to the Bootstrap},
	isbn = {978-0-429-24659-3},
	pagetotal = {456},
	publisher = {Chapman and Hall/{CRC}},
	author = {Efron, Bradley and Tibshirani, R. J.},
	date = {1994-05-15},
    year = {1994},
	doi = {10.1201/9780429246593},
}

@misc{brown_language_2020,
	title = {Language Models are Few-Shot Learners},
	url = {http://arxiv.org/abs/2005.14165},
	doi = {10.48550/arXiv.2005.14165},
	number = {{arXiv}:2005.14165},
	publisher = {{arXiv}},
	author = {Brown, Tom B. and Mann, Benjamin and Ryder, Nick and Subbiah, Melanie and Kaplan, Jared and Dhariwal, Prafulla and Neelakantan, Arvind and Shyam, Pranav and Sastry, Girish and Askell, Amanda and Agarwal, Sandhini and Herbert-Voss, Ariel and Krueger, Gretchen and Henighan, Tom and Child, Rewon and Ramesh, Aditya and Ziegler, Daniel M. and Wu, Jeffrey and Winter, Clemens and Hesse, Christopher and Chen, Mark and Sigler, Eric and Litwin, Mateusz and Gray, Scott and Chess, Benjamin and Clark, Jack and Berner, Christopher and {McCandlish}, Sam and Radford, Alec and Sutskever, Ilya and Amodei, Dario},
	urldate = {2025-05-07},
	date = {2020-07-22},
    year = {2020},
	eprinttype = {arxiv},
}

@misc{wei_chain--thought_2023,
	title = {Chain-of-Thought Prompting Elicits Reasoning in Large Language Models},
	url = {http://arxiv.org/abs/2201.11903},
	doi = {10.48550/arXiv.2201.11903},
	number = {{arXiv}:2201.11903},
	publisher = {{arXiv}},
	author = {Wei, Jason and Wang, Xuezhi and Schuurmans, Dale and Bosma, Maarten and Ichter, Brian and Xia, Fei and Chi, Ed and Le, Quoc and Zhou, Denny},
	urldate = {2025-05-07},
	date = {2023-01-10},
    year = {2023},
	eprinttype = {arxiv},
}

@misc{wu_large_2023,
	title = {Large Language Models are Diverse Role-Players for Summarization Evaluation},
	url = {http://arxiv.org/abs/2303.15078},
	doi = {10.48550/arXiv.2303.15078},
	number = {{arXiv}:2303.15078},
	publisher = {{arXiv}},
	author = {Wu, Ning and Gong, Ming and Shou, Linjun and Liang, Shining and Jiang, Daxin},
	urldate = {2025-05-07},
	date = {2023-09-19},
    year = {2023},
	eprinttype = {arxiv},
}

@misc{bai_longbench_2025,
	title = {{LongBench} v2: Towards Deeper Understanding and Reasoning on Realistic Long-context Multitasks},
	url = {http://arxiv.org/abs/2412.15204},
	doi = {10.48550/arXiv.2412.15204},
	shorttitle = {{LongBench} v2},
	number = {{arXiv}:2412.15204},
	publisher = {{arXiv}},
	author = {Bai, Yushi and Tu, Shangqing and Zhang, Jiajie and Peng, Hao and Wang, Xiaozhi and Lv, Xin and Cao, Shulin and Xu, Jiazheng and Hou, Lei and Dong, Yuxiao and Tang, Jie and Li, Juanzi},
	urldate = {2025-05-12},
	date = {2025-01-03},
    year = {2025},
	eprinttype = {arxiv},
}

@misc{li_long-context_2024,
	title = {Long-context {LLMs} Struggle with Long In-context Learning},
	url = {http://arxiv.org/abs/2404.02060},
	doi = {10.48550/arXiv.2404.02060},
	number = {{arXiv}:2404.02060},
	publisher = {{arXiv}},
	author = {Li, Tianle and Zhang, Ge and Do, Quy Duc and Yue, Xiang and Chen, Wenhu},
	urldate = {2025-05-12},
	date = {2024-06-12},
    year = {2024},
	eprinttype = {arxiv},
}

@article{trivedi2021musique,
  title={{M}u{S}i{Q}ue: Multihop Questions via Single-hop Question Composition},
  author={Trivedi, Harsh and Balasubramanian, Niranjan and Khot, Tushar and Sabharwal, Ashish},
  journal={Transactions of the Association for Computational Linguistics},
  year={2022},
  publisher={MIT Press}
}

@inproceedings{trotman_improvements_2014,
	location = {New York, {NY}, {USA}},
	title = {Improvements to {BM}25 and Language Models Examined},
	isbn = {978-1-4503-3000-8},
	url = {https://dl.acm.org/doi/10.1145/2682862.2682863},
	doi = {10.1145/2682862.2682863},
	series = {{ADCS} '14},
	pages = {58--65},
	booktitle = {Proceedings of the 19th Australasian Document Computing Symposium},
	publisher = {Association for Computing Machinery},
	author = {Trotman, Andrew and Puurula, Antti and Burgess, Blake},
	urldate = {2025-08-12},
    year = {2014},
	date = {2014-11-26},
}

@misc{noauthor_general_nodate,
	title = {General · lorenzbrehme/{CARE}},
	url = {https://github.com/lorenzbrehme/CARE},
author = {Lorenz Brehme, Thomas Ströhle, Ruth Breu},
	language = {en},
	urldate = {2026-02-17},
	journal = {GitHub},
}

@misc{moreira_enhancing_2024,
	title = {Enhancing Q\&A Text Retrieval with Ranking Models: Benchmarking, fine-tuning and deploying Rerankers for {RAG}},
	url = {http://arxiv.org/abs/2409.07691},
	doi = {10.48550/arXiv.2409.07691},
	shorttitle = {Enhancing Q\&A Text Retrieval with Ranking Models},
	number = {{arXiv}:2409.07691},
	publisher = {{arXiv}},
	author = {Moreira, Gabriel de Souza P. and Ak, Ronay and Schifferer, Benedikt and Xu, Mengyao and Osmulski, Radek and Oldridge, Even},
	urldate = {2024-11-06},
	date = {2024-09-12},
	eprinttype = {arxiv},
	eprint = {2409.07691},
}

@misc{xu_negative_2022,
	title = {Negative Sampling for Contrastive Representation Learning: A Review},
	url = {http://arxiv.org/abs/2206.00212},
	doi = {10.48550/arXiv.2206.00212},
	shorttitle = {Negative Sampling for Contrastive Representation Learning},
	number = {{arXiv}:2206.00212},
	publisher = {{arXiv}},
	author = {Xu, Lanling and Lian, Jianxun and Zhao, Wayne Xin and Gong, Ming and Shou, Linjun and Jiang, Daxin and Xie, Xing and Wen, Ji-Rong},
	urldate = {2025-08-15},
	date = {2022-06-01},
	eprinttype = {arxiv},
	eprint = {2206.00212 [cs]},
}

@misc{brehme_retrieval-augmented_2025,
	title = {Retrieval-Augmented Generation in Industry: An Interview Study on Use Cases, Requirements, Challenges, and Evaluation},
	url = {http://arxiv.org/abs/2508.14066},
	doi = {10.48550/arXiv.2508.14066},
	shorttitle = {Retrieval-Augmented Generation in Industry},
	number = {{arXiv}:2508.14066},
	publisher = {{arXiv}},
	author = {Brehme, Lorenz and Dornauer, Benedikt and Ströhle, Thomas and Ehrhart, Maximilian and Breu, Ruth},
	urldate = {2025-08-21},
	date = {2025-08-11},
	eprinttype = {arxiv},
	eprint = {2508.14066 [cs]},
}

\end{document}